# Layer sliding and twisting induced electronic transitions in correlated magnetic 1T-NbSe$_2$ bilayers

Jiaqi Dai[1, #], Jingsi Qiao[2, #], Cong Wang[1,#], Linwei Zhou[3,1], Xu Wu[2], Liwei Liu[2], Xuan Song[2], Fei Pang[1], Zhihai Cheng[1], Xianghua Kong[3], Yeliang Wang[2, *], and Wei Ji[1, *]

[1]*Beijing Key Laboratory of Optoelectronic Functional Materials & Micro-Nano Devices, Department of Physics, Renmin University of China, Beijing 100872, P. R. China*

[2]*MIIT Key Laboratory for Low-Dimensional Quantum Structure and Devices, School of Integrated Circuits and Electronics, Beijing Institute of Technology, Beijing 100081, P. R. China*

[3]*College of Physics and Optoelectronic Engineering, Shenzhen University, 3688 Nanhai Avenue, Nanshan District, Shenzhen, 518061, P. R. China*

[#] *These authors contributed equally to this work.*
*Correspondence and request for materials should be addressed to Y. W. (yeliang.wang@bit.edu.cn) and W.J. (wji@ruc.edu.cn)*

**Abstract:** Correlated two-dimensional (2D) layers, like 1T-phases of TaS$_2$, TaSe$_2$ and NbSe$_2$, exhibit rich tunability through varying interlayer couplings, which promotes the understanding of electron-correlation in the 2D limit. However, the coupling mechanism is, so far, poorly understood and was tentatively ascribed to interactions among the $d_{z^2}$ orbitals of Ta or Nb atoms. Here, we theoretically show that the interlayer hybridization and localization strength of interfacial Se $p_z$ orbitals, rather than Nb $d_{z^2}$ orbitals, govern the variation of electron-correlated properties upon interlayer sliding or twisting in correlated magnetic 1T-NbSe$_2$ bilayers. Each of the both layers is in a star-of-David (SOD) charge-density-wave phase. Geometric and electronic structures, and magnetic properties of 28 different stacking configurations were examined and analyzed using density-functional-theory calculations. We found that the SOD contains a localized region (Reg-L), in which interlayer Se $p_z$ hybridization plays a paramount role in varying the energy levels of the two Hubbard bands. These variations lead to three electronic transitions among four insulating states, which demonstrated the effectiveness of interlayer interactions to modulate correlated magnetic properties in a prototypical correlated magnetic insulator.

**Introduction**

Many exotic quantum phases, e.g. unconventional superconductor [1-3], quantum spin liquids [4,5], Mott and charge-transfer (CT) insulators [6-9], and Wigner crystals [10-12], have been emerging in correlated systems. In the strong-correlation limit of electron correlated systems, on-site Coulomb interactions split a half-filled band into two sub-bands, namely the lower (LHB) and upper Hubbard bands (UHB), forming a Mott or a CT insulator. [13] As the electronic screening substantially weakens in the two-dimensional (2D) limit, correlation effects are expected to play a more pronounced role in thin layered materials. 1H-phase layers of group-5 dichalcogenides, e.g. 1H-TaS$_2$ [14,15] and 1H-NbSe$_2$ [16-18], show electron correlation effects like charge density wave (CDW) [14,16] and superconductivity. [15,17,18] Their 1T-phase counterparts exhibit even diverse electron-correlated properties, including CDW [5,8,19], quantum spin liquid [5,20], Mott [21,22] or CT [7] insulating states at low temperatures.

Interlayer coupling was demonstrated to be a powerful route to substantially tune the correlated properties of these correlated 2D layers [23-28], although point defects could play a minor role [7,20]. Critical temperatures of CDW and superconductivity were enhanced in the H-phase monolayers by replacing an $h$-BN substrate [23] and layer intercalation [27], respectively. A Kondo lattice and a heavy fermion state emerge in artificially stacked 1T/1H and 1H/1T TaS$_2$ hetero-bilayers [24], respectively. In pure 1T-phase layers, the Mott gap is modulable by varying interlayer coupling strengths, as demonstrated in stacking induced Mott-insulator to metal [25] or doping induced Mott- to band-insulator transitions [26,28] in 1T-TaS$_2$. Such tunability was phenomenologically ascribed to interactions among Ta $d_{z^2}$ states [25,26,28], in which the interlayer S-S bonding mediate [25].

1T-NbSe$_2$ is expected to exhibit more fruitful properties upon layer stacking as interlayer out-of-plane Se $p_z$ states are involved. 1T-NbSe$_2$ is a CT insulator, in which the VB, comprised of the Se $p_z$ states, appreciably interferes with the LHB [7], different from the Mott insulating states in 1T-Ta-dichalcogenides [20,21,29,30]. These Se states are directly interlayer overlapped, rather than those $d_{z^2}$ states from the embedded and

~ 6 Å apart metal (Ta) atoms [25,26,28]. However, the 1T-NbSe$_2$ mono- and few-layers are, so far, experimentally accessible using molecular beam epitaxy solely [7,31-33]. This preparation difficulty is a roadblock to discovering the role of interlayer coupling in correlated properties of 1T-NbSe$_2$ few-layers. Surprisingly, the layer-stacking effects on correlated properties of 1T-NbSe$_2$ few-layers are still poorly understood theoretically, even though a preliminary experimental observation was reported during the writing stage of this work [33].

Here, we revealed the role of interlayer coupling in modulating Mott-Hubbard-related properties in √13 × √13 CDW bilayers of 1T-NbSe$_2$ using density functional theory calculations. In the CDW phase of the 1T-NbSe$_2$ monolayer, 13 NbSe$_2$ unit-cells in a star-of-David (SOD) (Fig. 1a) move towards the central Nb atom (Nb1), exhibiting a CT insulating states that the VB and UHB (Fig. 1e) states are separated by an insulating gap of $\varDelta$ = 0.42 eV (Fig. 1b and 1c). The interacting LHB (Fig. 1d) and VB states (Fig. 1c) result in a weak in-plane ferromagnetic (FM) spin-exchange coupling $J$ = 0.12 meV/SOD of the local magnetic moments around the Nb1Se1$_3$ clusters (Fig. 1a) [8,19] in the Heisenberg model (Fig. S1). We considered 28 layer-stacking configurations in total, including seven stacking sites, two twisting angles, and two types of interlayer magnetic couplings [FM and antiferromagnetic (AFM)]. By carefully analyzing their stacking-dependent electronic structures and magnetic properties, we found that the intra-layer FM coupling remains in 27 configurations where the in-plane Hubbard states persist. However, the interlayer coupling divides the bilayers into three major categories in terms of electronic structures. While Categories Cat-I and Cat-II include nonmagnetic (NM) band insulators and AFM CT insulators, a mixed CT and Mott insulator, showing the best thermal stability and interlayer FM, was found in Cat-III. The localization strength of the interlayer hybridized out-of-plane Se $p_z$ states was found to dominate those stacking-dependent electron-correlated properties in the 1T-NbSe$_2$ bilayer.

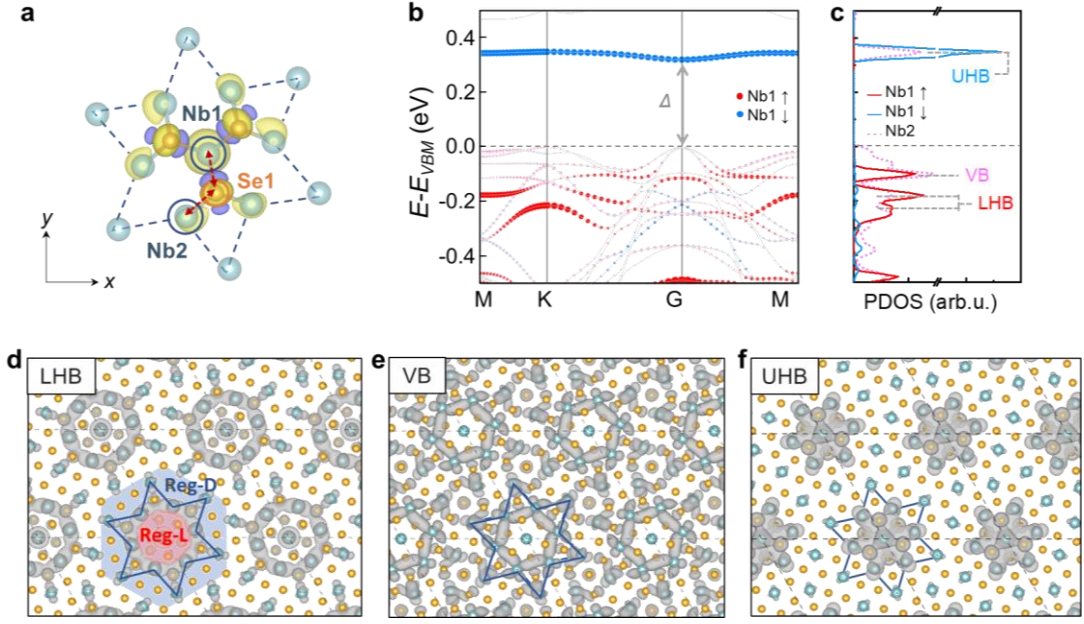

Figure 1. Geometric and electronic structures of monolayer NbSe$_2$. (**a**) Spin density in a SOD of monolayer 1T-NbSe$_2$. The red dashed arrows show the interacting Nb2-Se1-Nb1 atoms. Yellow and purple isosurface contours correspond to spin-up and -down charge densities, respectively. Blue (orange) balls represent Nb (Se) atoms. (**b-c**) Bandstructure (**b**) and corresponding density of states (**c**) projected on the spin-up (red) and -down (blue) components of the Nb1and the spin-averaged $d_{z^2}$ orbital of Nb2 (pink dashed line in **c**). LHB, VB, and UHB represent the lower Hubbard band, valence band, and upper Hubbard band, respectively. $\Delta$ denotes the bandgap. (d-e) Visualized wavefunction norms of LHB (**d**), VB (**e**), and UHB (**f**) at the Gamma point. The isosurface value was set to 3×10$^{-4}$ $e$/Bohr$^3$. Reg-L (red) and Reg-D (blue) marked in (*d*) represent the localized and delocalized regions, respectively.

**Results and Discussions**

**Stacking dependent geometry and magnetic order of 2L T-NbSe$_2$ CDW.** Figures 2a and 2b display all three Nb (N1–N3) and five Se (S1–S5) sites in the top- (Fig. 2a) and side-views (Fig. 2b) of a SOD in a 1T-NbSe$_2$ CDW monolayer, in which sites S3 and S4 are equivalent upon stacking in the bilayers. Site N3 is the farthest one from N1 among all sites inside the SOD (highlighted by dark blue dashed lines in Fig. 2a). Sites S3–S5 are outside the SOD and locate around N3. Figures 2c and 2d illustrate the two considered twisting angles of 0°and 60° upon stacking two 1T-NbSe$_2$ monolayers into a bilayer, which were evidenced in previous STM observations [31]. We used the N1 site of the bottom layer as the reference. Thus, the stacking configurations of the bilayers are notated like S2-R0-FM (Fig. 2e) or N2-R60-AFM (Fig. 2f). Here, N2 or S2 represents which site the top layer sits over the N1 site of the bottom layer, R60 (R0) indicates the twisting angle of 60° (0°), and FM or AFM denotes the interlayer magnetism.

Figure 2g plots the relative total energies of all these 28 stacking configurations as a function of the sliding path shown in Fig. 2a where the absolute total energy of N3-R60-FM was set to 0 eV. The interlayer FM state cannot be maintained in N1-R0, N1-R60, N2-R0, and S1-R60, thus their energies are not available in the plot. The configurations with the 60° twisting angle are generally more stable than their corresponding counterparts of 0°, except for sites N1 and S1 where the energies of the R0 configurations are slightly lower. The interlayer AFM coupling is overwhelmingly preferred in most configurations and the interlayer FM coupling is solely favored for the two N3 and the S5-R0 configurations. A qualitative trend of interlayer magnetism was displayed in Fig. 2h that sliding the top layer away from the AA stacking (N1-R0) generally stabilizes the interlayer FM coupling regardless of the twisted angle. Moreover, the amount of the stabilizing energy is positively related to the sliding distance that the larger the distance, the less significant or even reversed the FM-AFM energy difference.

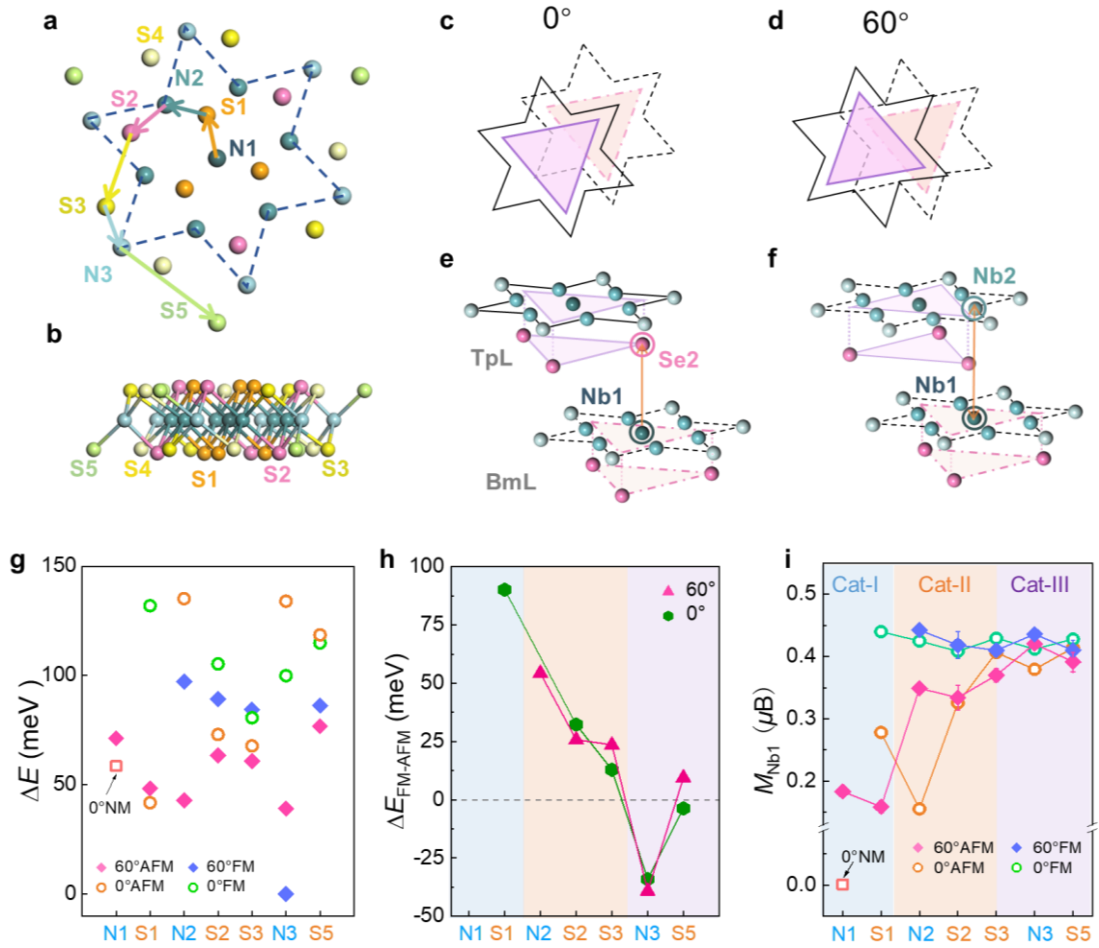

Figure 2. Notation, thermal stability, and magnetism of 28 stacking configurations of bilayer $NbSe_2$. (**a,b**) Schematic of the sliding path for seven sites of the 1T-$NbSe_2$ bilayer. The Nb1 site of the bottom layer (BmL) was set as the reference, and the path illustrates how the top layer (TpL) moves away from the AA stacking. Sites N1-N3 and S1-S5 represent the positions of three and five inequivalent Nb and Se atoms, respectively. Orange, pink, yellow, and green (Steel-, cadet-, and sky-blue) balls represent Se1, Se2, Se3, and Se5 (Nb1, Nb2, and Nb3) atoms, respectively (hereinafter). (**c-f**) Schematics of bilayer 1T-$NbSe_2$ stacked in twisting angles of 0° (**c,e**) and 60° (**d,f**) in the top (**c,d**) and side views (**e,f**). Bilayers S2-R0 and N2-R60 were illustrated in (**e**) and (**f**). (**g**) Total energies of the 28 different bilayers, which show interlayer non- (NM), ferro- (FM) and antiferro- magnetism(AFM). (**h**) Energy differences between the interlayer FM and AFM orders for the 14 stacking orders. (**i**) Local magnetic moments projected on Nb1 for the 28 bilayers. Cat-I, -II, and -III denote those three categories of the 1T-$NbSe_2$ bilayers.

This sliding distance also affects the size of the local magnetic moment ($M_{Nb1}$) projected on the Nb1 (the Nb atom sitting at the N1 site) atom. Figure 2i plots the $M_{Nb1}$ values of the all 28 configurations. $M_{Nb1}$ is approximately zero in the N1-R0-AFM bilayer, therefore, we denote it as N1-R0-NM for clarity. The moment remains small in preferred interlayer AFM configurations, including the N1-R60-AFM, S1-R$t$-X ($t$=0, 60 and X=AFM, FM), and N2-R0-AFM. Among these small $M_{Nb1}$ configurations, the N1 and S2 series configurations were categorized as Cat-I, given the appearance of their density of states (DOSs) for the LHB and UHB states as discussed later. The $M_{Nb1}$ value continuously increases when sliding the top layer further, which eventually reaches the saturation at ~ 0.4 $\mu_B$ in configurations S3 (Cat-II, including the N2, S2, S3, and S5-R60 series) and keeps stable in Cat-III configurations including the N3 and S5-R0 series. It is remarkable that the energy differences between the FM and AFM orders for S5-R60 (Cat-II) and S5-R0 (Cat-III) are less than 5 meV/SOD, but they exhibit different magnetic ground states. In comparison with the interlayer distances of 2.93 (3.45) Å for S5-R0-FM (N3-R60-FM), the shorter interlayer distance of 2.86 Å for S5-R60-AFM explains the more favored interlayer AFM coupling, as we discussed in the Bethe-Slater-curve like the behavior of the interlayer magnetism in TMDs (Fig. S2a and S2b) [34].

Such significant variation of interlayer distances ($d_1$ and $d_2$) is common among different configurations (Fig. S3). The N$n$-R0-X ($n$=1,2,3; X=NM, AFM, FM) series always have smaller interlayer distances than the corresponding 60-degree rotated ones, showing variations of the interlayer distance up to 0.6 Å between the AFM and FM configurations. The variation reduces to roughly 0.1 Å for the S series, in which the 60-degree-rotated configurations exhibit smaller interlayer distances. These significant variations could be intuitively understood that in all N$n$-R0-X configurations, an interfacial Se atom of the top (bottom) layer always sits at the hollow site of its adjacent Se atoms from the bottom (top) layer, effectively reducing the repulsion among them. The electronic hybridization of interfacial chalcogen atoms plays a paramount role in tuning electronic structures of TMD bi- and few-layers [34-36]. Given the varying interlayer Se-Se interactions, it could be inferred that diverse electronic properties are observable in different stacking configurations, as we elucidated below.

**Role of layer twisting in tuning electronic structures of NbSe$_2$ bilayer.** As we mentioned earlier that the interfacial hybridization of Se atoms plays a paramount role in tuning the electronic structures of bilayer 1T-NbSe$_2$. We noticed that the LHB and UHB states are distributed around the Nb1Se1$_3$ clusters solely. The Nb1 and Se1 atoms in the same cluster have the same majority spin-component, indicating that the total magnetization of the entire layer is locked with that of the interlayer Se1 atoms. Therefore, the stacking details of interfacial Se1 atoms, more precisely, the interfacial wavefunction overlaps (hybridizations) between Se1 and other Se atoms, primarily determine the electronic structure and magnetism of a given bilayer. We first considered two Cat-I configurations, namely N1-R0-NM and N1-R60-AFM, which are twisted by 60° to each other and represent the NM (Cat-I-A) and AFM (Cat-I-B) configurations found in Cat-I, highlighting the role of twisting in tuning electronic structures of the bilayer.

For the N1-R0-NM bilayer, all the three interfacial Se1 atoms of the top (TpL) [bottom (BmL)] layer reside at hollow sites of interfacial Se atoms from the TpL and BmL, showing a hexagonal pattern in the top-view (Fig. 3a). This hexagon-like stacking order substantially promotes interfacial wavefunction overlap among Se1 atoms. Each Se1 atom of the top (bottom) layer hybridizes with two Se1 and one Se2 atoms of the bottom (top) layer (Fig. 3a and 3b), forming interlayer "covalent-like quasi-bonds" (CLQBs) [34,37-43]. The originally localized LHB and UHB states hybridize into two energetically degenerate LHB (UHB) states through this interlayer interaction, which is extended to six interfacial Se2 atoms (Fig. 3a).

The bandstructures projected on Nb1 (Fig. 3c) appears to show a CT insulating state for N1-R0-NM, which was verified by the two pronounced sharp peaks in the DOS plot (Fig. 3d). However, those two peaks are spin-non-polarized ones, essentially different from the spin-polarized Mott peaks observable in the NbSe$_2$ monolayer. The spin-non-polarized LHB (UHB) is further verified that each of the two LHB (UHB) states is equally distributed in the top (Fig. 3e) and bottom (Fig. 3f) layers, which means these two LHBs are spin, energetically and spatially degenerated. In other words, substantial charge delocalization among the six Se1 and six Se2 atoms at the interfacial region leads to the bilayer showing an NM band, not a CT, insulating state.

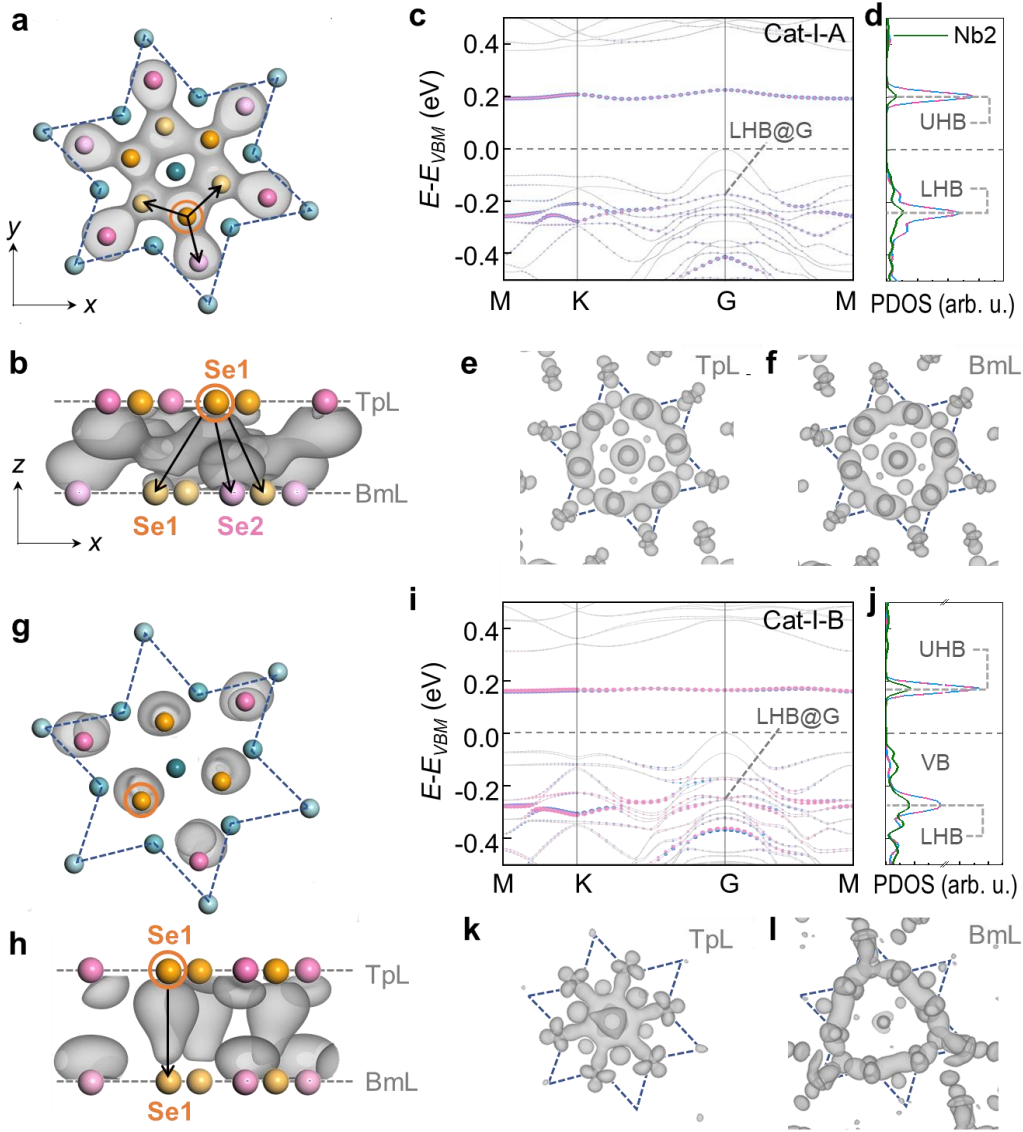

Figure 3. Interlayer couplings and electronic structures of Cat-I. (**a-b**) Visualized wavefunction norms of LHB at the Gamma point of N1-R0-NM (Cat-I-A) in the top (a) and side (b) views, which were clipped to show those of interfacial Se atoms solely. The isosurface value was set to $1\times10^{-5}$ $e$/Bohr$^3$. Darker and lighter orange (pink) balls represent Se1 (Se2) atoms in the top and bottom layers (TpL and BmL), respectively and blue balls denote Nb atoms. (**c-d**) Electronic bandstructure (c) and corresponding density of states (d) projected on the $d_{z^2}$ orbitals of the Nb1 atoms in the TpL (pink) and BmL (blue), and those of Nb2 atoms in the both layers (green). (**e-f**) layer-dependent wavefunction norms in TpL (**e**) and BmT (**f**) replotted from those shown in panels **a** and **b** but in a larger isosurface value of $3\times10^{-4}$ $e$/Bohr$^3$. (**g-l**) The same scheme of plots as that of panels **a**–**f** for the N1-R60-AFM bilayer.

The top layer of the N1-R60-AFM bilayer is twisted by 60° around the N1 site relative to that of the N1-R0-NM bilayer, which is the most stable bilayer in Cat-I and the third most stable one among all the 28 configurations. This rotation shifts the interfacial Se1 atoms of the top layer sitting right on-top of the interfacial Se1 atoms of the bottom layer (Fig. 3g). The triangular, rather than the hexagonal, configuration of the six interfacial Se1 atoms limits the interlayer wavefunction overlap to occur only between pairs of vertically aligned Se1 atoms, leaving isolated interlayer hybridized states (Figs. 3g and 3h). The much more localized interlayer hybridization keeps the spin-polarized Mott states but reduces $M_{Nb1}$ of the top and bottom layers. This on-top stacking configuration also leads to stronger interlayer Se1-Se1 repulsion, consistent with the approximately 0.6 Å larger interlayer distance. The remaining magnetic moments prefer to be oriented anti-parallel, showing an AFM CT insulator state with a 0.42 eV CT gap in the projected bandstructures of N1-R60-AFM (Fig. 3i). Like the N1-R0-NM case, there are two pronounced narrow peaks showing in the DOS plot (Fig. 3j), each of which is, differentiated from the N1-R0-NM case, comprised of two spin-polarized, anti-parallel oriented and layer dependent LHB and UHB states. Spatial distribution of the LHB (UHB) state (Fig. 3k and 3l) shows that it distributes at different regions of the top and bottom layers, consistent with the spin and spatially non-degenerate characteristic of the LHB (UHB). Thus, the AFM CT insulating state in the N1-R60-AFM bilayer represents other Cat-I configurations except N1-R0-NM.

**Role of layer sliding in tuning electronic structures of NbSe$_2$ bilayer.** Categories Cat-II and -III represent those configurations, in which the top-layers are further slid away from the S1 site. Two representative configurations, N2-R60-AFM of Cat-II (the fourth stable bilayer) and N3-R60-FM of Cat-III (the most stable bilayer), were plotted in Fig. 4, in which their magnetic ground states are different. Every Cat-II configuration has the interlayer AFM ground state. Although both the Cat-I-B and Cat-II configurations show the AFM ground state, the interlayer stacking order of Cat-II is substantially different from that of Cat-I-B where the localized Nb1Se1$_3$ clusters stack together. In each Cat-II configuration, the localized Nb1Se1$_3$ cluster, denoted Reg-L (see Fig. 1d), of the top (bottom) layer primarily resides over the delocalized region (Reg-D) of a SOD of the bottom (top) layer, forming an L-D interaction. Such interaction is illustrated by plots of differential charge densities (Fig. 4a). The spatially superposed Reg-L and Reg-D allow the LHB and VB states in the vicinity energy level to hybridize into several 'LHB-like' states (Fig. 4b), which smears the original LHB showing multiple wide peaks in the DOS plot (Fig. 4c). The UHB peak remains pronounced and sharp with a tiny energy splitting (Fig. 4c) that is resulted from the minorly superposed Reg-Ls. Thus, it could be regarded as a delocalized CT (dCT) insulator. The weakening interlayer CLQB and reducing Se1-Se1 Pauli repulsion in Cat-II configurations enlarge the magnetization of Nb1 (Fig. 2i) and maintain the interlayer AFM magnetic ground state.

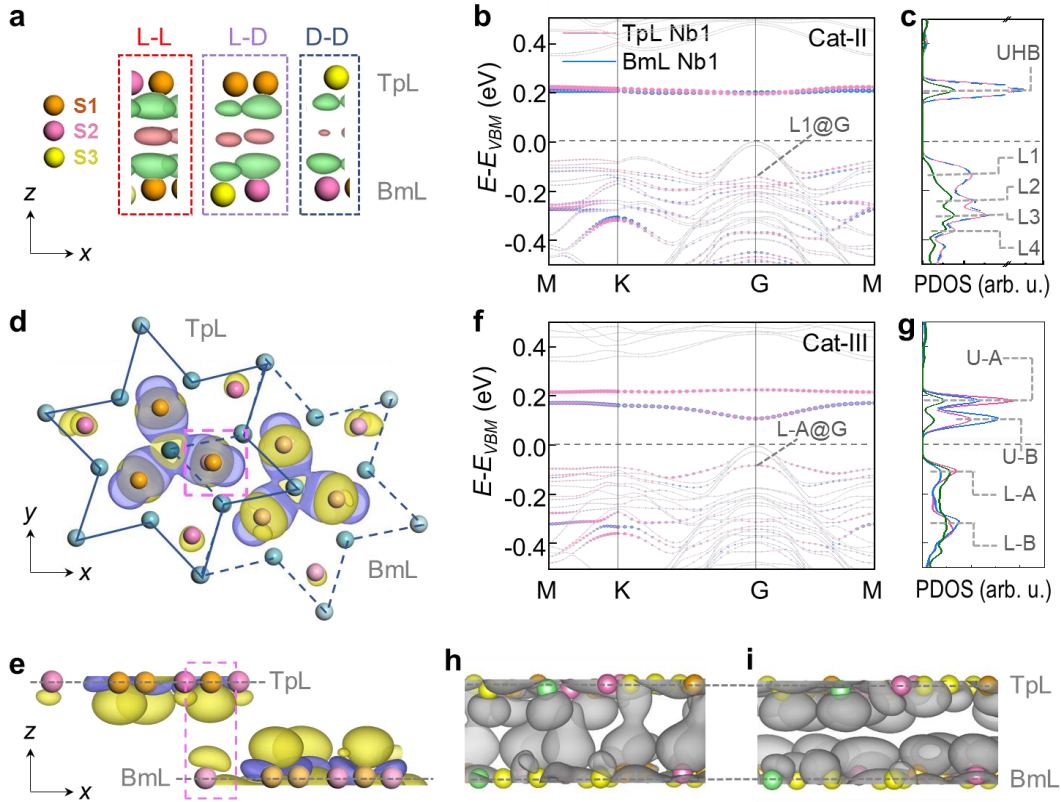

Figure 4. Interlayer couplings and electronic structures of Cat-II and -III. (**a**) Interlayer differential charge densities plotted among interfacial Se atoms using an isosurface value of $3\times10^{-4}$ $e$/Bohr$^3$. Red (green) isosurface contours represent charge accumulation (reduction). Notations L-L, L-D, and D-D represent the interactions between the two Reg-Ls of the both layers, Reg-L and Reg-D, and the two Reg-Ds of the both layers. (**b,c**) Electronic bandstructure (b) and corresponding density of states (c) projected on the $d_{z^2}$ orbitals of the Nb1 atoms in the TpL (pink) and BmL (blue), and of Nb2 atoms in the both layers (green). Four 'LHB-like' states

were identified and labeled as L1–L4. (**d,e**) Spin density of N3-R60-FM where yellow and purple isosurface contours correspond to spin-up and -down charge densities, respectively. The isosurface value was $3\times10^{-4}$ $e$/Bohr$^3$. The pink dashed rectangle highlights the interacting interfacial Se1 and Se2 atoms. (**f,g**) Electronic bandstructure (f) and corresponding density of states (g) are plotted in the same scheme as those for panels **b** and **c**. Bonding and anti-bonding UHB (LHB) states were labeled as U-B and U-A (L-B and L-A), respectively. (**h-i**) Visualized wavefunction norms of the L-B and L-A states at the Gamma point in an isosurface value of $5\times10^{-5}$ $e$/Bohr$^3$. Densities plotted in panels **a**, **d**, **e**, **h**, and **i** were clipped to show those between interfacial Se atoms solely

In Cat-III, Reg-L of the top layer fully superposes with Reg-D of the bottom layer, eliminating the L-L interaction, which substantially reduces the Pauli repulsion between interfacial Se atoms because of electron delocalization in Reg-D. Figures 4d-4e show the spin density of N3-R60-FM. A weak, but appreciable, interlayer electron hopping occurs between the same spin component of top-layer Se1 and bottom-layer Se2 atoms, as outlined by the red dashed boxes. This FM hopping, together with largely reduced Se-Se Pauli repulsion, favors the interlayer FM coupling in Cat-III bilayers, in which the $M_{Nb1}$ value is comparable to that of the monolayer. The interlayer hybridization of the same spin component splits the two LHB (UHB) states by 0.20 (0.11) eV at the Gamma point (Fig. 4h and 4i), forming a pair of interlayer bonding [L-B (U-B)] and anti-bonding [L-A (U-A)] states for LHB (UHB) (Fig. 4f and 4g) and a correlated bandgap ($E_{UHB}$-$E_{LHB}$) of 0.21 eV reduced from 0.43 eV. For UHBs, the interlayer bonding state bridges the Reg-Ls of the top and bottom layers and leads to U-B being more dispersive and in a lower energy. (Fig. S4) The splitting of LHB promotes the L-A to be the highest VB outside the pocket around the G point, showing a Mott insulating state, although it remains a CT insulator within the G-pocket.

**Interlayer interactions between electron localized and delocalized regions.** Given the four representative cases of all 28 bilayers, we summarized and schematically illustrated the roles of interlayer interaction variations in inducing transitions among band-, CT-, and Mott-insulators. We emphasized that the interfacial Se $p_z$ orbitals, not the Nb $d_{z^2}$ orbitals, play a dominant role rather than a mediating role in tuning the correlated properties in 1T-NbS$_2$ bilayers (Fig. 5). A comparably dominant role was played by interfacial I $p$ orbitals, rather than Cr $d$ orbitals, in changing the interlayer magnetism of CrI$_3$ bilayers. [44-47] We recall the definition of Reg-L, which refers to a spin-polarized and electron localized region observable around the Nb1Se1$_3$ cluster in either LHB or UHB, while Reg-D is assigned to the rest of the layer that is spin-non-polarized and electron delocalized. The bilayer could have interlayer L-L, L-D, and D-D interactions. In Cat-I, the L-L interaction (Se1-Se1 hybridization) dominates the magnetism solely (Fig. 5a and 5b). The hybridized states keep less dispersive and show two narrow peaks in the DOS plots, although the peaks are of different characteristics. In N1-R0-NM of Cat-I-A, the LHB and UHB peaks are spin-non-polarized owing to delocalized states formed within Reg-L through interfacial hybridization of six Se1 and six Se2 atoms (Fig 5a), which eliminate the local magnetic moments and lead the bilayer to a band-insulator.

Twisting by 60° (N1-R60-AFM) or a small sliding (S1-R0-AFM), however, breakdowns the delocalized hybridization and limits hybridized state distrusting within the superposed Se1-Se1 pairs (Fig. 5b). This change of interlayer hybridization preserves the local magnetic moments although their sizes are reduced by over half. In these cases, the bilayer is a CT insulator. The interlayer L-L interaction (Se1-Se1 hybridization) further weakens in Cat-II, and the L-D interaction raises (Fig. 5c). In comparison with a typical CT insulator, its LHB is strongly mixed up with interlayer VB states that obstruct experimental identification of its exact position in e.g. scanning tunneling spectroscopy. [33] As we termed it earlier, it is a dCT insulator. In the far limit of sliding, the direct L-L interaction eliminates and the D-D interaction governs in Cat-III bilayers (Fig. 5d). The two Mott states, which could be equivalently considered as two parallelly aligned magnetic moments, from the two layers indirectly interact through delocalized VB states (primarily the Se2 $p_z$ orbitals). This indirect interaction prefers the FM ground state and splits both the LHB and the UHB, showing a mixing feature of Mott and CT insulators.

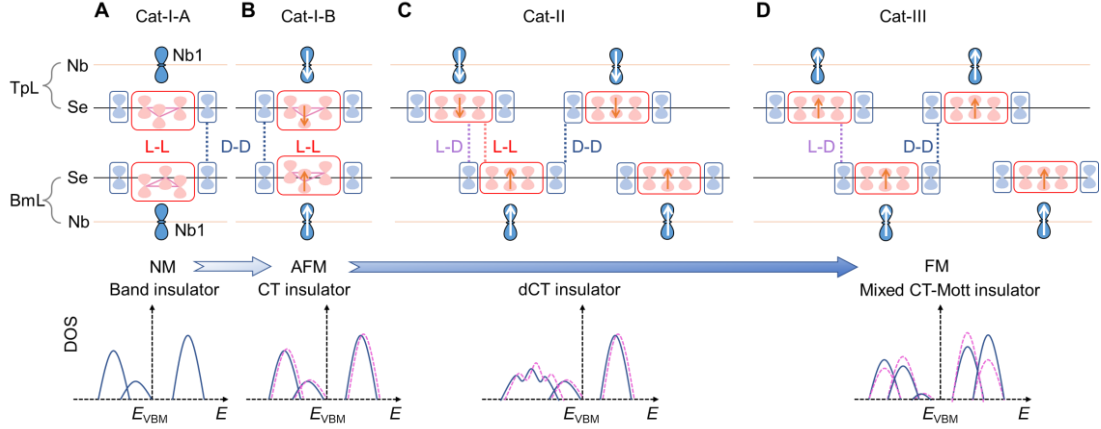

Figure 5. Illustration of interlayer interactions and magnetism for Cat-I-A (**a**), Cat-I-B (**b**), Cat-II (**c**), and Cat-III (**d**). The $d_{z^2}$ orbital of Nb (steel blue dumbbells)) and the pz orbitals of interfacial Se in Reg-L (red dumbbells) and Reg-D (sky blue dumbbells). White and orange arrows on Nb1 and Se1 illustrate the directions of magnetic moments, respectively. Red, purple, and blue dashed lines represent the interactions of the L-L, L-D, and D-D interlayer coupling channels, respectively. The arrows at the bottom illustrate the evolution of electronic structure and/or interlayer magnetism.

**Conclusion**

In short, we carried out density functional theory calculations on different stacking configurations of 1T-NbSe$_2$ bilayers to reveal the role of interlayer stacking in modulating their electron-correlated properties. The magnetic orders and the transitions among four insulating states are effectively tunable by constructing the overlapping (interacting) modes of Reg-L (where the localized Mott electron resides) and Reg-D (where delocalized valence electrons distribute) in the 1T-NbSe$_2$ bilayers. The variations of the L-L, L-D, and D-D interactions, changed by interlayer sliding or twisting, induce the band-to-CT (twisting with the L-L interaction), CT-to-dCT (introducing the L-D interaction), and dCT-Mott (eliminating the L-L interaction) insulating transitions. The varied interlayer interactions also change the Mott or CT gap. All these striking results highlight the importance of interlayer coupling in tunning correlated electronic states in NbSe$_2$ bilayers. The coexistence of localized and delocalized regions is often observed in several CDW layers [26,28]. We thus expect this 'overlapping regions' modulating mechanism could be a general strategy for constructing novel correlated electronic states in those CDW layers, which enables us to explore their properties that are potentially useful in future spintronic and quantum devices.

## METHODS

**DFT calculations.** Density functional theory calculations were performed using the generalized gradient approximation for the exchange-correlation potential, the projector augmented wave method [48,49], and a plane-wave basis set as implemented in the Vienna *ab-initio* simulation package (VASP) [50]. On-site Coulomb interactions were considered on the Nb *d* orbitals with an effective value $U$ = 3.2 eV. [51] Van der Waals interactions were considered at the DFT-D3 [50] level with the Perdew-Burke-Ernzerhof (PBE) functional.[50] The standard PBE functional was used with consideration of spin-orbit coupling (SOC) to calculate energies and electronic properties of all considered magnetic configurations, except for the single layer case where we intended to show the LHB and UHB are originated from different spin components. A kinetic energy cut-off of 500 eV was used for the plane-wave basis set for geometric and electronic structure calculations. A *k*-mesh of 7×7×1 was adopted to sample the first Brillouin zone of the SOD of the BL NbSe$_2$ in all calculations. The mesh density of *k* points was kept fixed when calculating the properties for different stacking configurations. All atoms in the supercell were allowed to relax until the residual force per atom was less than $1\times10^{-4}$ eV Å$^{-1}$ and the energy convergence criteria was $1\times10^{-5}$ eV. In plotting DOS spectra, a Gaussian (first-order Methfessel-Paxton) smearing of 0.02 eV was used for all bilayer configurations (the monolayer). The energy level of VBM was set to energy zero in all DOS and bandstructure calculations.

## SUPPLEMENTARY MATERAILS

Supplementary material for this article is available at http://xxx.

# REFERENCES

[1] Y. Cao, V. Fatemi, S. Fang, K. Watanabe, T. Taniguchi, E. Kaxiras, and P. Jarillo-Herrero, Nature **556**, 43 (2018).

[2] J. M. Park, Y. Cao, K. Watanabe, T. Taniguchi, and P. Jarillo-Herrero, Nature **590**, 249 (2021).

[3] P. A. Lee, N. Nagaosa, and X.-G. Wen, Reviews of Modern Physics **78**, 17 (2006).

[4] M. Klanjšek, A. Zorko, R. Žitko, J. Mravlje, Z. Jagličić, Pabitra K. Biswas, P. Prelovšek, D. Mihailovic, and D. Arčon, Nat. Phys. **13**, 1130 (2017).

[5] K. T. Law and P. A. Lee, Proceedings of the National Academy of Sciences **114**, 6996 (2017).

[6] Y. D. Wang, W. L. Yao, Z. M. Xin, T. T. Han, Z. G. Wang, L. Chen, C. Cai, Y. Li, and Y. Zhang, Nat. Commun. **11**, 4215 (2020).

[7] M. Liu *et al.*, Science Advances **7**, eabi6339 (2021).

[8] Y. Chen *et al.*, Nat. Phys. **16**, 218 (2020).

[9] Y. Cao *et al.*, Nature **556**, 80 (2018).

[10] E. C. Regan *et al.*, Nature **579**, 359 (2020).

[11] Y. Zhou *et al.*, Nature **595**, 48 (2021).

[12] H. Li *et al.*, Nature **597**, 650 (2021).

[13] J. Zaanen, G. A. Sawatzky, and J. W. Allen, Phys. Rev. Lett. **55**, 418 (1985).

[14] R. Grasset, Y. Gallais, A. Sacuto, M. Cazayous, S. Mañas-Valero, E. Coronado, and M.-A. Méasson, Phys. Rev. Lett. **122**, 127001 (2019).

[15] E. Navarro-Moratalla *et al.*, Nat. Commun. **7**, 11043 (2016).

[16] C.-S. Lian, C. Si, and W. Duan, Nano Lett. **18**, 2924 (2018).

[17] X. Xi, Z. Wang, W. Zhao, J.-H. Park, K. T. Law, H. Berger, L. Forró, J. Shan, and K. F. Mak, Nat. Phys. **12**, 139 (2016).

[18] M. M. Ugeda *et al.*, Nat. Phys. **12**, 92 (2016).

[19] L. Liu *et al.*, Nat. Commun. **12**, 1978 (2021).

[20] W. Ruan *et al.*, Nat. Phys. **17**, 1154 (2021).

[21] Y. Fei, Z. Wu, W. Zhang, and Y. Yin, AAPPS Bulletin **32**, 1 (2022).

[22] D. Shin, N. Tancogne-Dejean, J. Zhang, M. S. Okyay, A. Rubio, and N. Park, Phys. Rev. Lett. **126**, 196406 (2021).

[23] W. Fu *et al.*, ACS Nano **14**, 3917 (2020).

[24] W. Wan, R. Harsh, A. Meninno, P. Dreher, S. Sajan, I. Errea, F. de Juan, and M. M. Ugeda, arXiv e-prints, arXiv:2207.00096 (2022).

[25] S.-H. Lee, J. S. Goh, and D. Cho, Phys. Rev. Lett. **122**, 106404 (2019).

[26] C. Butler, M. Yoshida, T. Hanaguri, and Y. Iwasa, Nat. Commun. **11**, 1 (2020).

[27] H. Zhang *et al.*, Nat. Phys. **18**, 1425 (2022).

[28] J. Lee, K.-H. Jin, and H. W. Yeom, Phys. Rev. Lett. **126**, 196405 (2021).

[29] W. Zhang *et al.*, Phys. Rev. B **105**, 035110 (2022).

[30] Z. Wu *et al.*, Phys. Rev. B **105**, 035109 (2022).

[31] Y. Chen, L. Liu, X. Song, H. Yang, Z. Huang, T. Zhang, H. Yang, H.-J. Gao, and Y. Wang, 2D Materials **9**, 014007 (2022).

[32] Q. Zhang *et al.*, ACS Nano **15**, 16589 (2021).

[33] Liwei Liu *et al.*, ACS Nano, DOI: 10.1021/acsnano.2c10841 (2023).

[34] C. Wang, X. Zhou, L. Zhou, Y. Pan, Z.-Y. Lu, X. Wan, X. Wang, and W. Ji, Phys. Rev. B **102**, 020402 (2020).

[35] Y. Zhao, J. Qiao, P. Yu, Z. Hu, Z. Lin, S. P. Lau, Z. Liu, W. Ji, and Y. Chai, Advanced Materials **28**, 2399 (2016).

[36] Y. Zhao *et al.*, Advanced Materials **29**, 1604230 (2017).

[37] J. Qiao, Y. Pan, F. Yang, C. Wang, Y. Chai, and W. Ji, Science Bulletin **63**, 159 (2018).

[38] Z.-X. Hu, X. Kong, J. Qiao, B. Normand, and W. Ji, Nanoscale **8**, 2740 (2016).

[39] J. Qiao, X. Kong, Z.-X. Hu, F. Yang, and W. Ji, Nat. Commun. **5**, 4475 (2014).

[40] C. Wang, X. Zhou, Y. Pan, J. Qiao, X. Kong, C.-C. Kaun, and W. Ji, Phys. Rev. B **97**, 245409 (2018).

[41] B. Li *et al.*, Nat. Mater. **20**, 818 (2021).

[42] Y. Huang *et al.*, Nat. Commun. **11**, 2453 (2020).

[43] Q. Fu *et al.*, Adv. Sci. **9**, 2204247 (2022).

[44] P. Jiang, L. Li, Z. Liao, Y. Zhao, and Z. Zhong, Nano Lett. **18**, 3844 (2018).

[45] P. Jiang, C. Wang, D. Chen, Z. Zhong, Z. Yuan, Z.-Y. Lu, and W. Ji, Phys. Rev. B **99**, 144401 (2019).

[46] D. Soriano, C. Cardoso, and J. Fernández-Rossier, Solid State Commun. **299**, 113662 (2019).

[47] S. W. Jang, M. Y. Jeong, H. Yoon, S. Ryee, and M. J. Han, Phys. Rev. Mater. **3**, 031001 (2019).

[48] P. E. Blöchl, Phys. Rev. B **50**, 17953 (1994).

[49] G. Kresse and D. Joubert, Phys. Rev. B **59**, 1758 (1999).

[50] G. Kresse and J. Furthmüller, Phys. Rev. B **54**, 11169 (1996).

[51] M. Cococcioni, Correlated Electrons: From Models to Materials Modeling and Simulation **2** (2012).

## Acknowledgements

We gratefully acknowledge financial support from the Ministry of Science and Technology (MOST) of China (Grant No. 2018YFE0202700), the National Natural Science Foundation of China (Grants No. 11974422, 92163206, 12204534, 61888102, 61971035 and 62171035), the Strategic Priority Research Program of the Chinese Academy of Sciences (Grant No. XDB30000000), the Beijing Nova Program from Beijing Municipal Science & Technology Commission (Z211100002121072), the Fundamental Research Funds for the Central Universities, and the Research Funds of Renmin University of China (Grants No. 22XNKJ30). N.L. is grateful to the China Postdoctoral Science Foundation (2022M713447) for partially financial support. Calculations were performed at the Physics Lab of High-Performance Computing of Renmin University of China and the Shanghai Supercomputer Center.

## Author Contributions

W.J., Y.W and J.Q conceived this research. J.D, J.Q., C.W and W.J. performed the calculations and theoretical analysis. J.D, J.Q., and W.J. wrote the manuscript and all the authors commented on it.

## Competing interests

The authors declare no competing financial interests.

## Data and materials availability

All data needed to evaluate the conclusions in the paper are present in the paper and/or the Supplementary Materials. Additional data related to this paper may be requested from the authors. Correspondence and request for materials should be addressed W.J. (wji@ruc.edu.cn) and Y. W. (yeliang.wang@bit.edu.cn).

# Supporting Information

# Layer sliding and twisting induced electronic transitions in correlated magnetic 1T-NbSe$_2$ bilayers

Jiaqi Dai[1, #], Jingsi Qiao[2, #], Cong Wang[1,#], Linwei Zhou[3,1], Xu Wu[2], Liwei Liu[2], Xuan Song[2], Fei Pang[1], Zhihai Cheng[1], Xianghua Kong[3], Yeliang Wang[2, *], and Wei Ji[1, *]

[1]*Beijing Key Laboratory of Optoelectronic Functional Materials & Micro-Nano Devices, Department of Physics, Renmin University of China, Beijing 100872, P. R. China*

[2]*MIIT Key Laboratory for Low-Dimensional Quantum Structure and Devices, School of Integrated Circuits and Electronics, Beijing Institute of Technology, Beijing 100081, P. R. China*

[3]*College of Physics and Optoelectronic Engineering, Shenzhen University, 3688 Nanhai Avenue, Nanshan District, Shenzhen, 518061, P. R. China*

[#] *These authors contributed equally to this work.*

*Correspondence and request for materials should be addressed to Y. W. (yeliang.wang@bit.edu.cn) and W.J. (wji@ruc.edu.cn)*

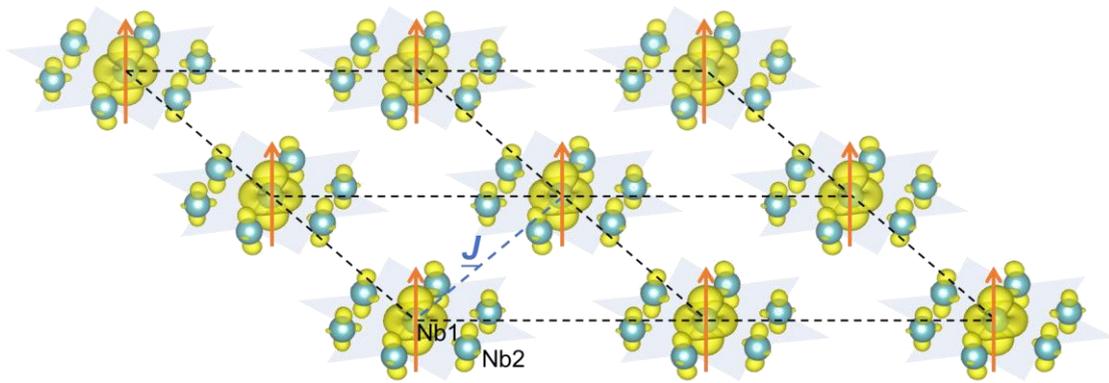

**Figure S1. The spin density of monolayer NbSe₂.** The orange arrows represent the magnetic moments in SOD. The Blue dashed line illustrates in-plane *J*.

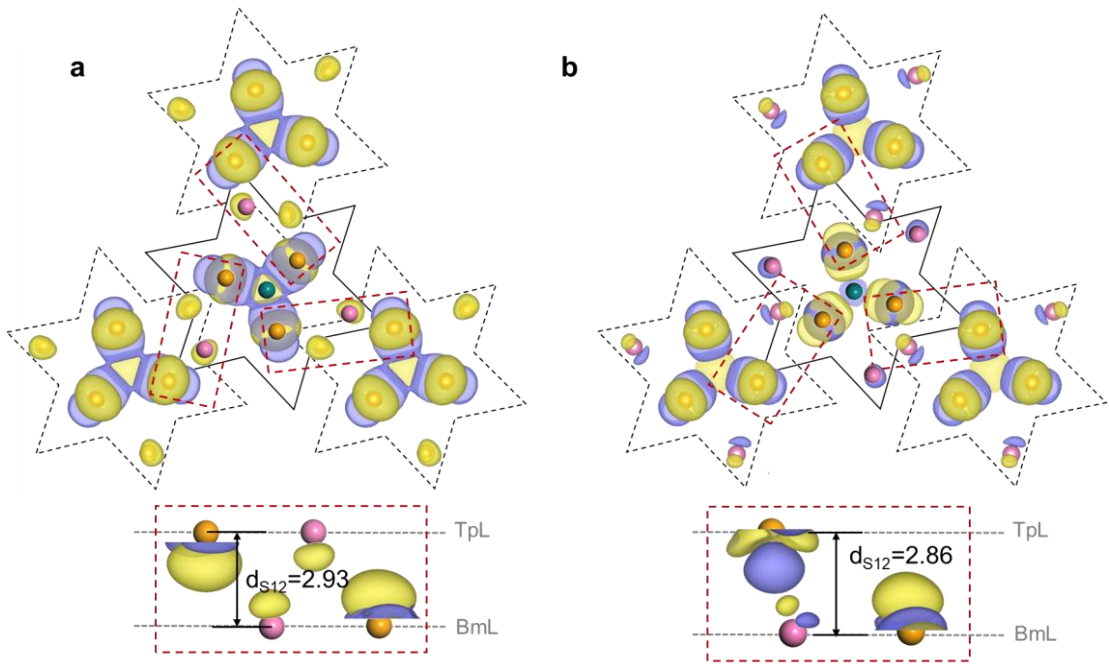

**Figure S2. The spin density of S5-R0-FM (a) and S5-R60-AFM (b).** Yellow and purple isosurface contours correspond to spin-up and -down charge densities, respectively. The red dotted rectangle shows the interacting interfacial Se1 and Se2 atoms. $d_1$ represents the vertical interlayer distance.

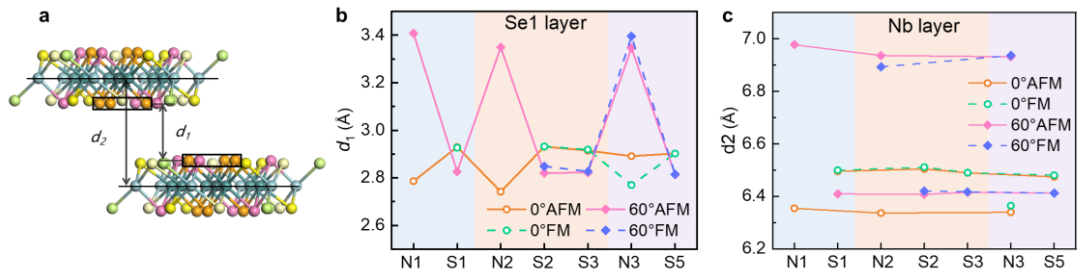

**Figure S3. Geometric properties of BL-NbSe$_2$.** (a) Atomic structure of BL-NbSe$_2$. $d_1$ represents the average vertical interlayer distance of interfacial Se atoms marked by a black solid rectangle. $d_2$ represent the average vertical interlayer distance of two Nb layers marked by the black solid line. (b-c) $d_1$ and $d_2$ of all these 28 stacking configurations as a function of the sliding path shown in Fig. 2a.

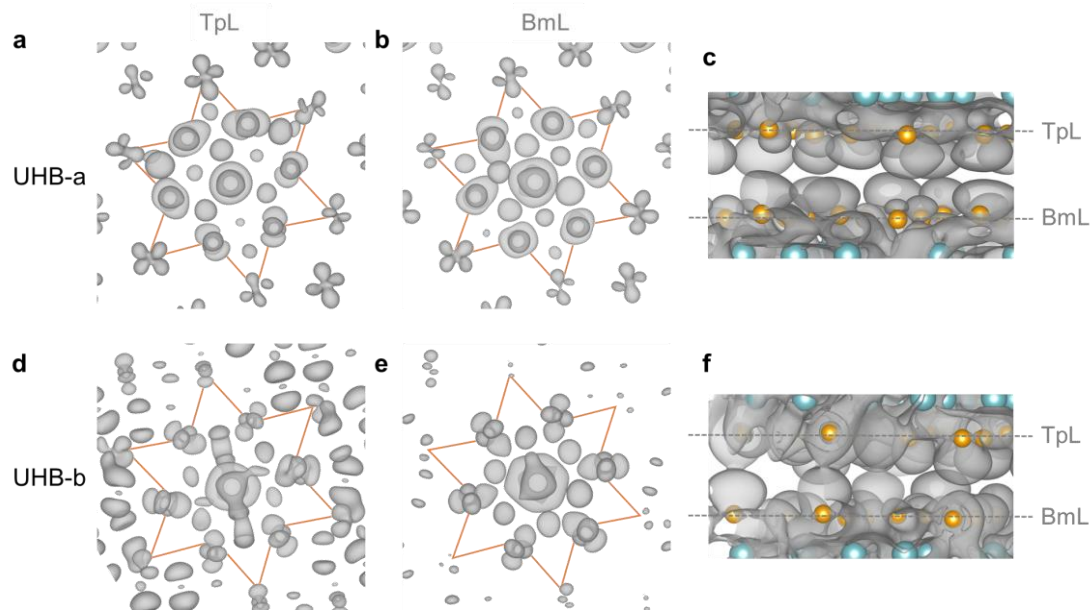

**Figure S4. Visualized wavefunction norms of BL-NbSe$_2$.** Visualized wavefunction norms of UHB at the Gamma point for the top layer (a, d), bottom layer (b, e) and interlayer (c, f) of N3-R60-FM. UHB-a and UHB-b represent the antibonding and bonding UHB states.